\documentclass[aps,pre,twocolumn,showpacs,superscriptaddress,groupedaddress,10pt]{revtex4-1}  

\usepackage{graphicx}  
\usepackage{grffile}
\usepackage{dcolumn}   
\usepackage{bm}        
\usepackage{amssymb}   
\usepackage{amsmath}
\usepackage{amsthm}
\usepackage{amsthm}
\usepackage{mathtools}
\usepackage{xcolor}
\usepackage{tikz}	
\usepackage{ulem}	
\usepackage[ansinew]{inputenc}
\usepackage{calrsfs}

\begin{document}



\title{Changeover phenomenon in randomly colored Potts models}

\date{\today}

\author{Nir Schreiber}
\affiliation{Department of Mathematics, Bar Ilan University, Ramat Gan, Israel 5290002}
\author{Reuven Cohen}
\affiliation{Department of Mathematics, Bar Ilan University, Ramat Gan, Israel 5290002}
\author{Gideon Amir}
\affiliation{Department of Mathematics, Bar Ilan University, Ramat Gan, Israel 5290002}
\author{Simi Haber}
\affiliation{Department of Mathematics, Bar Ilan University, Ramat Gan, Israel 5290002}

\begin{abstract}
A hybrid Potts model where a random concentration $p$ of the spins
assume $q_0$ states and a random concentration $1-p$ of the spins assume $q>q_0$ states is introduced.
It is known that when the system is homogeneous, with an integer spin number $q_0$ or
$q$, it undergoes a second or a first order transition, respectively.
It is argued that there is a concentration $p^\ast$ such that
the transition nature of the model is changed at $p^\ast$.
This idea is demonstrated analytically and by simulations
for two different types of interaction:
 the usual square lattice nearest neighboring and mean field all-to-all.
Exact expressions for the second order critical line in concentration-temperature parameter space of the mean field model together with some other related critical properties, are derived.
\end{abstract}

\maketitle
\section{Introduction}
\label{sec:intro}
The Potts model \cite{Potts1952,Wu1982} has been extensively studied in the past few decades
and the amount of related literature is enormous.
Though the model has some experimental realizations, it is mostly useful as a
laboratory for analytical
concepts and simulations methods \cite{Baxter1973,Nienhuis1979,Nauenberg1980,Cardy1980,Mukamel1976,Wilson1989,
Honmura1984,Feldmann1998,Graner1992,Bayong1999,Binder1981,Swendsen1987,Ferrenberg1988,Wang2001,Wang2001a}.
In the context of temperature-driven phase transitions,
the Potts model can be realized as a generalization to the celebrated Ising model
with the typical ``up-down" symmetry,
as it has a multi-fold symmetry that is spontaneously broken.

An interesting problem that has been first studied by Baxter \cite{Baxter1973}
is the dependence of the transition nature (first versus second order) on the spin number $q$.
 Baxter considered the square lattice ferromagnetic model with nearest neighbors interaction
(henceforth referred to as the \textit{standard model})
and obtained an exact expression for the latent heat.
That expression remained finite for $q>4$ and vanished for $q=4$.
Using an equivalence of the Potts model to a six-vertex model \cite{Temperley1971}
Baxter has shown that the free energy of the latter near the critical temperature
was negligibly small. Based on these findings, Baxter conjectured that the model in subject
exhibits a continuous transition for $q\leq q_c$ and discontinuous transition
for $q>q_c$ where $q_c=4$ is the \textit{changeover} integer or the maximal integer
for which the transition is continuous.
Recently, Duminil-Copin \textit{et al} \cite{DuminilCopin2016,1611.09877} have rigorously
proven Baxter's conjecture using the random cluster or the Fortuin-Kasteleyn 
\cite{Fortuin1972} representation of the Potts model.
Another interesting manifestation of $q_c=4$ being a changeover integer
is the presence of a multiplicative logarithmic correction term to the leading order power law 
divergence at criticality of, e.g., the specific heat and the magnetic susceptibility
\cite{Cardy1980,Salas1997}. This is in contrast to $q<4$ where this term is absent.

Though the changeover integer $q_c=4$ has been found for the square lattice, it is 
expected, due to universality, to hold 
for other translation-invariant lattices \cite{Nienhuis1979}.
More precisely,  
two dimensional systems accompanying the spontaneous breaking of the $q$-fold Potts symmetry, are expected to maintain the changeover behavior of the standard model for a wide range of local interaction patterns. There are, however, a few counterexamples \cite{Schreiber2019,Tamura2010,Tanaka2011}.

In \cite{Tamura2010,Tanaka2011} the authors introduced the so-called 
Potts model with \textit{invisible colors} (IC)
where the spins assume $q$ ``visible" colors (Potts states) as in the 
\textit{local interaction} standard model and additional $r$ ``invisible" colors
that control the entropy of but do not affect the energy of the system.
The model has a growing interest in recent years and problems such as its marginal dimensions \cite{Krasnytska2016} or its behavior on general Bethe lattices \cite{Ananikian2013},
thin graphs \cite{Johnston2013} or scale-free networks \cite{Sarkanych2019}, have been addressed.

It has been demonstrated \cite{Tamura2010} using a mean field solution to the Blume-Emery-Griffiths model \cite{Blume1971}
together with simulations, and
also rigorously proven \cite{vanEnter2011,VANENTER2012}, that the IC model
exhibits a first order transition for any $q$ when $r$ is sufficiently large.
A non-trivial outcome of the latter result is the occurrence of
a first order transition for $q\leq 4$.
The reason is that, typically, in the continuous transition cases large fractal-like configurations
buffer between order and disorder, and, even though the number of such fractals grows exponentially with their size, it does not compensate the entropic cost of their construction when
the total number of microstates is large enough.

We present a hybrid Potts (HP) model that captures the concept of manipulating the
transition order, while keeping the interaction parameters fixed, by introducing inhomogeneity in the number of colors. Strictly speaking, we
consider a Potts Hamiltonian ${\cal H}(\{\sigma\})$
\footnote{Any Potts Hamiltonian known to describe a system undergoing
a ferromagnetic phase transition may apply.}
where $\{\sigma\}$ is a configuration
of $N$ spins, each can select one out of the colors $\{1,2,...,Q_i\}$  where
\begin{equation}
\label{eq:spins}
Q_i = -X_i(q-q_0)+q,\ i=1,...,N\;,
\end{equation}
are \textit{i.i.d.} random variables with $X_i\sim {\rm{Ber}}(p)$. In other words, each spin can be colored in $q_0\leq q_c$ ``strong" colors with probability $p$ and in $q>q_c$ colors (containing the same $q_0$ colors and additional  $q-q_0$
 ``weak" colors), with probability $1-p$.
We say that spins chosen with probability $p$ or $1-p$ belong to strong
or weak regions, respectively.

Since the HP model
assumes varying concentrations of spins with different types of colors, between limiting homogeneous instances
that lead to a transition which is either continuous or discontinuous, one might expect that at some concentration, $0<p^\ast<1$, the model displays a \textit{changeover phenomenon},
that is, switches from one type of transition to another for every
combination of $q_0\leq q_c$ strong colors and a total number of $q>q_c$ colors.

The rest of the paper is organized as follows. In section \ref{sec:2d} we find $p^\ast$
for the standard model. 
We also present the results of Monte-Carlo (MC) simulations.
A detailed analysis of the mean field (MF) model
equipped with the HP prescription \eqref{eq:spins} is reported in section \ref{sec:MF}.
Finally, our observations are summarized and discussed in section \ref{sec:summary}.

\section{The standard model}
\label{sec:2d}
In the present section we detect $p^\ast$ using a simple
energy-entropy first principle argument.
To be more specific, we compute the entropy loss
against the energy gain in generating long range order (LRO),
where it is known that in a second order transition,
LRO is usually established by monochromatically coloring fractal clusters that form an exponential family with positive geometric entropy, and in a first order transition
LRO is associated with the formation of monochromatic clusters that are taken from a sub-exponential 
family having zero geometric entropy. By comparing the changes in the corresponding free energies, 
$p^\ast$ can be found.

We start with a system of $N$ spins situated on the vertices of the square lattice and interacting via the usual Hamiltonian
\begin{equation}
\label{eq:H_B}
{\cal H}_B = -J\sum_{\langle i,j\rangle}\delta_{\sigma_i,\sigma_j}\;,
\end{equation}
where $J > 0$ is the ferromagnetic coupling constant ($J$, together with $k_B$, is henceforth set to one for convenience) and the summation is taken over nearest neigboring spins.

We call a large cluster \textit{simple} if its number of sites per bond
is minimal, i.e., $1/2 + o(1)$.
We term a \textit{snake}, a fractal with a maximal number of $1+o(1)$ sites per bond.
Any fractal therefore grows no faster than a snake and has $1/2+\delta + o(1)$
sites per bond, where $\delta$ is a positive parameter no larger than $1/2$.

Consider a fractal contour made of $l = O(N)$ bonds.
The contour has energy $-l$. Suppose the fractal has $n$ sites where   
$k$ of them are positioned in strong regions and $n/l = 1/2 + \delta + o(1)$. Let $\#(k,n)$ be the number
of such fractals. The
expected number of these fractals is
\begin{equation}
\label{eq:kn}
\langle \# (k,n)\rangle = \mu ^n {n\choose k}p^k(1-p)^{n-k}\;,
\end{equation}
where $\sum_ k \#(k,n) \sim \mu^n$ and the growth number $\mu$ in
general depends on $\delta$ \cite{Schreiber2019}. In making such fractals monochromatic, 
the entropy of the system is changed in the amount of
$\ln \left(\# (k,n) q_0^{-k}q^{-(n-k)}\right)$. Assuming 
the quantity $\#(k,n)$ is narrowly distributed around the mean (NDAM) and
using \eqref{eq:kn},
the change in the free energy per site with inverse temperature $\beta$, can be written
(to leading order in $N$)
\begin{eqnarray}
\label{eq:snake}
& &-\beta \Delta f_{\rm{frac}}  = \frac{2\beta}{1+2\delta} + \ln\mu -\ln q\\
& &+\sum_k\Bigg( \frac{1}{n}\ln \left({n\choose k}p^k(1-p)^{n-k}\right)+
\frac{k}{n}\ln\left(\frac{q}{q_0}\right)\Bigg )\nonumber\;.
\end{eqnarray}
Under the assumption of NDAM, the sum in \eqref{eq:snake} can be replaced with the
maximum of the summand obtained at $k$ satisfying
\begin{equation}
\label{eq:k_max}
\kappa = \frac{pq}{pq+(1-p)q_0}\;,
\end{equation}
with $\kappa = k/n$. This brings \eqref{eq:snake} into the form
\begin{eqnarray}
\label{eq:snake_max}
-\beta \Delta f_{\rm{frac}} &=& \frac{2\beta}{1+2\delta} + \ln\mu -\ln q - \kappa\ln \kappa \nonumber\\
&-& (1-\kappa)\ln (1-\kappa) + \kappa\ln p \nonumber\\ 
&+&(1-\kappa )\ln(1-p)+\kappa\ln\left(\frac{q}{q_0}\right) \;.
\end{eqnarray}
Since, unlike fractals, large simple clusters grow sub-exponentially
with their size, they are randomly distributed across the lattice. Accordingly,
the change in the free energy per site due to simple clusters with the same energy $-l$ is given by
\begin{eqnarray}
\label{eq:DelS}
-\beta \Delta f_{\rm{sim}}= 2\beta - p\ln q_0 - (1-p)\ln {q}\;.
\end{eqnarray}

It may be more constructive to form large monochromatic simple clusters
rather than fractals. Specifically, within the framework of \eqref{eq:snake_max},\eqref{eq:DelS},
if at $\beta$ solving $\Delta f_{\rm{sim}}=0$
we have $\Delta f_{\rm{frac}}\geq 0$, then it is entropically disadvantageous for the system to possess
large fractals at that temperature. Instead, large simple monochromatic clusters are formed
and  the system undergoes a first order transition \cite{Schreiber2019}
at
\begin{equation}
\label{eq:Tc1st_approx}
\beta_c \approx \frac{1}{2}\left( p\ln q_0 + (1-p)\ln {q}\right)\;.
\end{equation}
The marginal concentration $p^\ast$ below which large simple clusters are entropically
more favorable than fractals can be estimated by
taking $\kappa^\ast$ to satisfy \eqref{eq:k_max} at $p^\ast$,
plugging it together with the RHS of \eqref{eq:Tc1st_approx} 
into \eqref{eq:snake_max} and solving
\begin{eqnarray}
\label{eq:p0*}
\sup_{\delta} &\Bigg(& \frac{p^\ast\ln q_0 + (1-p^\ast)\ln q}{1+2\delta} + 
\ln\mu - \ln q \nonumber\\
& -&\kappa^\ast\ln \kappa^\ast
- (1-\kappa^\ast)\ln (1-\kappa^\ast)  + \kappa^\ast\ln p^\ast \nonumber\\ 
&+& (1-\kappa^\ast )\ln(1-p^\ast) 
 +  \kappa^\ast\ln\left(\frac{q}{q_0}\right)\Bigg) = 0
\end{eqnarray}
for $p^\ast$.
It may be useful to write \eqref{eq:p0*} as 
\begin{equation}
\label{eq:p*}
\sup_\delta \varphi(\delta,p^\ast,q_0,q) = 0\;.
\end{equation}
Apart from uncovering the dependence of $p^\ast$ on 
the spin numbers $q_0,q$, \eqref{eq:p*} tells that $p^\ast$ is indeed marginal,
since, at a temperature that is candidate to be the first order critical point,  
$\Delta f_{\rm{frac}} = 0$, even for fractals that are
taken from the most probable exponential family.

For a large total number of colors, a first order transition occurs when the concentration of strong regions is small.
The reason is that, for $q$ large, the construction of monochromatic exponential families of fractals is entropically too expensive when the system is too sparsely populated by 
$q_0$-type spins.
The critical behavior in this case is determined by the simple clusters
and the transition is of first order. 
In order to quantify how sparse the strong regions are,
it may be useful to find $p^\ast$ in the large $q$ limit. To this end
we expand $\varphi(\delta,p^\ast,q_0,q) $ up to $o(p^\ast)$ and then solve \eqref{eq:p*} for $p^\ast$,
keeping the leading order large $q$ term. This results in a monotonically decreasing
behavior of $p^\ast$. In particular, if for varying $q$ the sup in \eqref{eq:p*} is 
determined by values of $\delta$ that are not arbitrarily small, i.e., satisfying $\inf_q\delta>0$, we have
\begin{equation}
\label{eq:p*_small}
p^\ast\sim \frac{\ln q}{q}\;,
\end{equation}
which means that a first order transition takes place for small concentrations,
typically below $\frac{\ln q}{q}$, of strong regions.

It should be noted that even when the concentration of strong spins is well below the percolation threshold, the presence of these spins can affect the transition nature.
Specifically, when $q$ is large, sparsely populated strong spins 
may lead to a second order transition since they considerably reduce the entropic cost of forming a monochromatic cluster having one of the strong colors.

Our next goal is to numerically support the theory presented up to this point. For that purpose, we
performed extensive MC simulations using the Wang-Landau (WL)
method \cite{Wang2001,Wang2001a} 
to calculate the density of states with energy $E$, $\Omega(E)$. 
Being equipped with the quantities $\Omega(E)$ 
enables us to compute the energy probability-distribution function (PDF)
at any desired temperature. In particular, we are interested in the PDF 
at pseudo-criticality, e.g., at a temperature
where the specific heat $C_L = L^{-d}\beta^2(\langle E^2\rangle - \langle E\rangle^2)$
with the energy moments given by
$\langle E^n\rangle = \frac{\sum_E E^n \Omega(E)e^{-\beta E}}{\sum_E \Omega(E)e^{-\beta E}}$,
is maximal \cite{Schreiber2018,Schreiber2019}.
The pseudo-critical (PC) temperatures, $T_L$, together with the maximal values
of the specific heat for different values of $p$, for a sample with 
a linear size $L = 40$ ($L = \sqrt N$) and spin numbers $q_0=2,\ q = 50$, are given in Table \ref{tab:table1}.
In order to roughly estimate the error bars of $T_L$ 
we ran the WL simulations $10$ times for a concentration $p=1$ and
$L = 40$. Taking the std of
the PC temperatures obtained from these runs,
where the temperature resolution in the canonical ensemble is 0.0001, gives
$0.0008$, thus, a plausible estimate of the error bars would be $0.001$. 
The error bars of the maxima are computed using the jackknife method.
\begin{table}[hbt!]
\caption{\label{tab:table1}
PC temperatures, $T_L$, and maximal values of the specific heat, $\max C_L$,
for different concentrations.
The spin numbers are $q_0=2,\ q=50$ and the lattice linear size is $L=40$.
The uniform error bar ($0.001$) of the temperatures is roughly estimated using $10$ WL runs.
The jackknife method is employed to compute the error bars of $\max C_L$.}
\begin{ruledtabular}
\begin{tabular}{lcl}
$p$              & $T_L$ & $\max C_L$\\
\hline
0 & 0.4793 & 3989$\pm$ 16 \\ 
0.05  & 0.4977 & 3405$\pm$  9\\ 
0.1  & 0.5132 & 173.29$\pm$ 0.01 \\
0.15 & 0.5316 & 61.498$\pm$ 0.001 \\
\end{tabular}
\end{ruledtabular}
\end{table}

\begin{figure}[hbt!]
\begin{center}
\includegraphics[width = 1.\columnwidth]{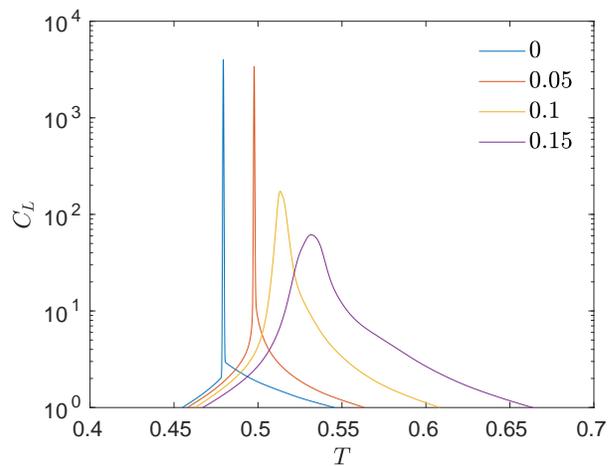}
\caption{Specific heat as a function of temperature, for a sample with a linear size $L=40$,
and spin numbers $q_0=2,\ q=50$, displayed on a semilogarithmic scale.
The peaks are positioned at temperatures given in Table \ref{tab:table1}.
Note the differences in the peaks' magnitude between 
the two lowest concentrations and the other concentrations.
}
\label{fig:C_LQ02Q50}
\end{center}
\end{figure}

In Fig.\ \ref{fig:C_LQ02Q50}, a plot of the specific heat against 
temperature is given for different 
concentrations. It can be verified that the peaks for the concentrations $p=0$ and $p=0.05$
are similar in magnitude and width.
On the other hand, the peaks of the larger concentrations
are (at least) an order of magnitude smaller and also broader.
This may qualitatively indicate different scaling behaviors and 
thus different types of transitions.

\begin{figure}[hbt!]
\begin{center}
\includegraphics[width = 1.\columnwidth]{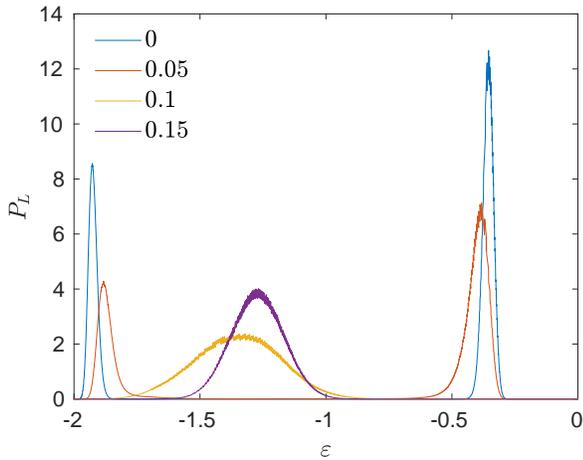}
\caption{PC energy PDF for the bond-interaction model on the square lattice.
The ``strong" Potts variable is $q_0=2$ and the total number of colors is $q=50$.
The lattice linear size is $L=40$.}
\label{fig:PDFQ02Q25}
\end{center}
\end{figure}

In Fig.\ \ref{fig:PDFQ02Q25} we plot the PC energy PDF
against the energy (per site).
The temperatures at which the PDFs are computed are those presented in Table \ref{tab:table1}.
For $p=0$, the values of the ordered and disordered energies (the positions of the two peaks) 
are $\varepsilon_o \simeq -1.9275$ and $\varepsilon_d \simeq -0.3544 $, respectively,
yielding $-(\varepsilon_o + \varepsilon_d) \simeq 2.2819$, 
in excelent agreement with $-(\varepsilon_o + \varepsilon_d) = 2(1 + 1/\sqrt {50}) = 2.282842...$ \cite{Janke1993}.
The excellent agreement between the PC temperature and 
the exact $T_c = 1/\ln(1 + \sqrt{50})= 0.478862...$ \cite{Hintermann1978}
should be also noted.
Furthermore, since $L=40$ is larger than the correlation
length \cite{Buddenoir1993}, it is likely that the
typical first order scaling law $|T_L - T_c| = O(L^{-d})$ 
($|T_L - T_c|\simeq 0.0004$, c.f. $40^{-2} \simeq  0.0006$), is satisfied.

In a second order transition, the PC energy PDF is expected to display
a single peak centred around the PC energy.
On the other hand, 
in a first order transition, a double peaked distribution centred around the energies of
the coexisting ordered and disordered states is conventional \cite{Janke1993}.
Evidently, while the broad PDF for $p=0.1$ is a manifestation of either a weak first order transition
or a single second order peak that suffers from finite size effects,
two pronounced peaks are observed for concentrations lower than $p=0.1$
and a clear single peak is present for $p=0.15$. The marginal concentration
 for a first order transition is therefore expected to be $0.05 < p^\ast \leq 0.15$, in reasonable agreement with \eqref{eq:p*_small} yielding $p^\ast\sim \frac{\ln 50}{50}\simeq 0.08$.

\section{Mean field model}
\label{sec:MF}
In the present section we adopt the Bragg-Williams approach to solve the  MF HP model.
The MF Hamiltonian is given by
\begin{equation}
\label{eq:H_MF}
{\cal H}_{MF} = -\frac{1}{N}\sum_{i<j}\delta_{\sigma_i,\sigma_j}\;,
\end{equation}
where the normalization factor $N^{-1}$ assures that the total energy is extensive.
Let $\xi_i$ and $\eta_i$ be the fraction of spins with color $i\in\{0,1,...,q_0-1,...,q-1\}$,
in strong and weak regions, respectively.
The total energy can be expressed as
${\cal E} = -\frac{1}{2}N\sum_i (p\xi_i+(1-p)\eta_i)^2$ and the number of states with this energy is given by
$W = {pN\choose {pN\xi_0,...,pN\xi_{q-1}}}{(1-p)N\choose {(1-p)N\eta_0,...,(1-p)N\eta_{q-1}}}$.
Thus, introducing the Lagrange multipliers $a,b$, the free energy per site takes the form
\begin{eqnarray}
\label{eq:betaf}
& &\beta f = \lim_{N\to\infty}\frac{1}{N}(\beta{\cal E}-\ln W) + 
{\mathrm{constraints}}  \nonumber\\
&=&\sum_i \Big(p\xi_i\ln \xi_i + (1-p)\eta_i\ln \eta_i
-\frac{1}{2}\beta (p\xi_i+(1-p)\eta_i)^2\Big)\nonumber \\
&+& a\Big(\sum_i \xi_i-1\Big)+b\Big(\sum_i \eta_i-1\Big)\;.
\end{eqnarray}
Setting $\xi_j=0,\ j=q_0,...q-1$,
differentiating \eqref{eq:betaf} with respect to $\xi_j,\eta_j$ and equating to zero give
\begin{eqnarray}
\ln\xi_j - \beta (p\xi_j+(1-p)\eta_j)& &\nonumber\\
+ a/p+1 = 0\;& &, j = 0,...,q_0-1\;,\nonumber\\
\ln\eta_j - \beta (p\xi_j+(1-p)\eta_j)& & \nonumber\\
+ b/(1-p)+1 = 0\;& &, j = 0,...,q-1 \;.
\end{eqnarray}
The fractions $\xi_j,\eta_j$ therefore satisfy the coupled equations
\begin{equation}
\label{eq:xi_exact}
        \xi_j=
        \begin{cases}
e^{\beta(p\xi_j + (1-p)\eta_j)}/Z_0 &,j = 0,...,q_0-1\\
0 &, j = q_0,...,q-1\;,\\
        \end{cases}
\end{equation}
and
\begin{equation}
\label{eq:eta_exact}
        \eta_j=
        \begin{cases}
e^{\beta(p\xi_j + (1-p)\eta_j)}/Z &,j = 0,...,q_0-1\\
e^{\beta(1-p)\eta_j}/Z &,j = q_0,...,q-1\;,\\
        \end{cases}
\end{equation}
where $Z_0 = e^{\frac{a}{p}+1}=\sum_{i=0}^{q_0-1}e^{\beta(p\xi_i + (1-p)\eta_i)}$
and $Z = e^{\frac{b}{1-p}+1}=Z_0 + \sum_{i=q_0}^{q-1}e^{\beta(1-p)\eta_i}$
are the colors partition functions in strong and weak regions, respectively.
The quantities in \eqref{eq:xi_exact},\eqref{eq:eta_exact} can take the form
\begin{equation}
\label{eq:xi}
        \xi_j=
        \begin{cases}
\frac{1}{q_0}\left(1 + (q_0-1)m_0\right) &, j=0\\
\frac{1}{q_0}(1-m_0) &,j = 1,...,q_0-1\\
0 &, j = q_0,...,q-1\;,\\
        \end{cases}
\end{equation}
and
\begin{equation}
\label{eq:eta}
        \eta_j=
        \begin{cases}
\frac{1}{q}(1+(q_0-1)m_0)\\
\times (1 + (q/q_0-1)m_1)   &, j=0\\
\frac{1}{q}(1-m_0) (1 + (q/q_0-1)m_1)  &,j = 1,...,q_0-1\;\\
\frac{1}{q}\left(1-m_1\right)  &,j = q_0,...,q-1\;,
        \end{cases}
\end{equation}
where $m_0,m_1$ are the components of a two-fold magnetization $\boldsymbol m$. Indeed,
there are favored fractions $\xi_0\geq \xi_j$ and $\eta_0 \geq \eta_j$
for all $1\leq j\leq q-1$, assuring the existence of LRO.
Eqs. \eqref{eq:xi},\eqref{eq:eta} also make sure that the ratio between the favored $j=0$ color and the other $q_0-1$ strong colors in both regions is preserved.

Above the critical point, at the disordered state, strong colors are uniformly distributed so that
\eqref{eq:xi},\eqref{eq:eta} imply
\begin{equation}
\label{eq:eq_weight}
m_0 = 0\;.
\end{equation}
In the case that the transition is continuous,
when plugging \eqref{eq:xi},\eqref{eq:eta} into \eqref{eq:betaf},
the magnetization at the critical point, $\boldsymbol{m}^\ast$,
must satisfy $\nabla_{\boldsymbol m} f \equiv (g_0,g_1)=0$ 
 where \eqref{eq:eq_weight} holds.
It is known \cite{Wu1982,Kihara1954} that for
the homogeneous model $q_c=2$.
Thus, in the following we set $q_0=2$.
The gradient components then read
\begin{widetext}
\begin{eqnarray}
g_0 & = &
-\frac{1}{2 q^2}\Bigg((1-p) q \left(2 + m_1 (q-2)\right)\ln \left(\frac{1-m_0}{1+m_0}\right)
     +\beta m_0 (m_1 (1-p) (q-2)+p (q-2)+2)^2\Bigg)
     + p \tanh ^{-1}(m_0)\nonumber \;,\label{eq:g0}  \\ \\
g_1 & = &
-\frac{(1-p) (q-2)}{2 q^2} \Bigg(\beta \left(1+m_0^2\right) q (m_1 (1-p)+p)
 +  2 \beta m_0^2 (1-m_1) (1-p)
 +  q \left(m_0 \ln\left(\frac{1-m_0}{1+m_0}\right)\right.\nonumber\\
 &-&  \left.2\ln\left(\frac{2+m_1(q-2)}{2q}\right)-\ln\left(1-m_0^2\right)\right)\Bigg)
 \label{eq:g1}\;.
\end{eqnarray}
\end{widetext}
Substituting \eqref{eq:eq_weight} in \eqref{eq:g0}, $g_0$ indeed vanishes.
The critical temperature is obtained by the further condition
 that the magnetization at criticality is an unstable point of the free energy.
 This can be formulated by considering the 
  Hessian matrix $H(\boldsymbol{m},\beta,p,q)$ given by

\begin{widetext}
\begin{equation}
\left(
\begin{array}{cc}
 -\frac{(m_1 (1-p) (q-2)-p (q-2)-2) \left(\beta (1-m_0^2) (q p-2 p+m_1 (1-p) (q-2)+2)-2 q\right)}{2 \left(1-m_0^2\right) q^2} & \frac{(1-p) (q-2) \left(2 \beta m_0 (-q p+2 p-m_1 (1-p) (q-2)-2)-
 q \ln \left(\frac{1-m_0}{1+m_0}\right)\right)}{2 q^2} \\
\frac{(1-p) (q-2) \left(2 \beta m_0 (-q p+2 p-m_1 (1-p) (q-2)-2)-
 q \ln \left(\frac{1-m_0}{1+m_0}\right)\right)}{2 q^2} & -\frac{(1-p) (q-2)
 \left(\beta (1-p) ((q-2) m_0^2+q)-\frac{2 q^2}{(1-m_1) (m_1 (q-2)+2)}\right)}{2 q^2}\\
\end{array}
\right)\;,
\end{equation}
\end{widetext} 
computed at $\boldsymbol{m}^\ast$ satisfying
$m_0^\ast=0,\ g_1(\boldsymbol{m}^\ast,\beta_c,p,q)=0$, where the Hessian has a vanishing eigenvalue, or, equivalently, obeys
\begin{equation}
\label{eq:detH}
\det H(\boldsymbol{m}^\ast(\beta_c,p,q),\beta_c,p,q)=0\;.
\end{equation}
Fixing $q$,
\eqref{eq:detH} implicitly determines the second order critical line in concentration-temperature plane by
\begin{equation}
\label{eq:Tc2nd_exact}
\frac{1}{2}\exp{\left(\frac{\beta_c-q}{q-2}\right)} = \frac{\beta_c-2}{(q-2)(2-\beta_c p)}\;.
\end{equation}
The critical magnetization reads
\begin{equation}
\label{eq:m_c}
\boldsymbol{m}^\ast = \left(0,\frac{q(2-\beta_c p)}{\beta_c(1-p)(q-2)}-\frac{2}{q-2}\right)\;.
\end{equation}
Plugging \eqref{eq:m_c} into the energy part of \eqref{eq:betaf}
yields the following expression for the critical energy
\begin{equation}
\label{eq:e_c}
\varepsilon_c = -\frac{(\beta_c-4) \beta_c+2 q}{2 \beta_c^2 (q-2)}\;.
\end{equation}

It is expected 
that the critical temperature, $\beta_c$, behaves as a continuous function
 of the concentration $p\in [0,1]$
\footnote{Indeed, the implicit functions theorem guarantees that, for
  $q$ fixed, \eqref{eq:Tc2nd_exact} defines a
  continuously differentiable function $\beta _c = \psi (p),\ p\in(0,1)$.}. However, the solution to \eqref{eq:Tc2nd_exact} at $p=0$ gives $\beta_c=q$, whereas the homogeneous system undergoes
  a first order transition at $\beta_c = \frac{2(q-1)\ln(q-1)}{q-2}$ \cite{Wu1982}.
Thus, continuity implies that there exists a finite concentration $p^\ast$ where for $p<p^\ast$ \eqref{eq:Tc2nd_exact} does not hold, and the transition becomes a first order transition. It should be noted that $p^\ast$ is expected to vary with $q$.

A first order transition is associated with a discontinuity of
the magnetization at the critical point.
In other words, there are two points $\boldsymbol{m}^\ast$
and $\boldsymbol{m}^{\ast\ast}$ that simultaneously minimize the free energy, that is, simultaneously solve $\nabla_{\boldsymbol m} f = 0,\ f(\boldsymbol{m}^\ast) = f(\boldsymbol{m}^{\ast\ast})$.

\begin{figure}[hbt!]
\begin{center}
\includegraphics[width = 1.\columnwidth]
{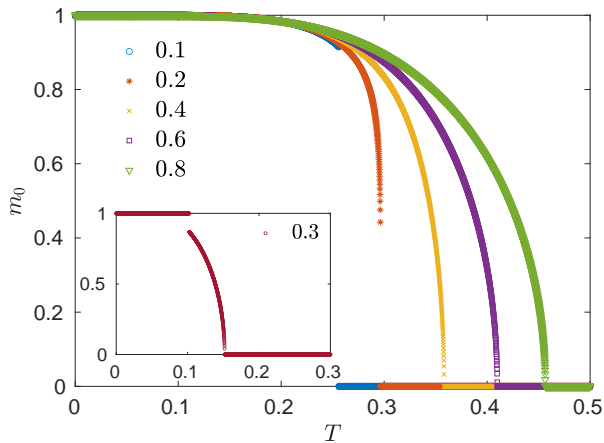}
\caption{Strong magnetization component $m_0$ as a function of temperature,
for different concentrations. The total number of colors is $q=6$.
It should be noted that the symbols composing the graphs
for $p\geq 0.4$ are more dense than for $p\leq 0.2$, in some neighborhood of $T_0$.
A combination of continuous and discontinuous
transitions at critical temperatures near $T_0=0.1518$ and $T_0=0.1021$, respectively,
for $q=1000$ and $p=0.3$, is presented in the inset.
}
\label{fig:m0_minimizing_f}
\end{center}
\end{figure}

The nature of the transition and, consequently, the presence of $p^\ast$,
can be phenomenologically detected by fixing the concentration and
examining the behavior of $\boldsymbol{m}$ that numerically minimizes \eqref{eq:betaf}
when the temperature is varied.
In Fig.\ \ref{fig:m0_minimizing_f}, the ``strong" component $m_0$
is plotted against temperature, for $q=6$ and different concentrations. Apparently, for $p\leq 0.2$ the magnetization has a discontinuity in the vicinity of $T_0$
(taken to be the point where the numerical derivative $\frac{\Delta m_0}{\Delta T}$
is ``large"),
indicating the presence of $m_0^\ast=0,m_0^{\ast\ast}>0$
\footnote{In practice we measure, e.g., for $p=0.1$, $m_0(0.2560) = 0.9149$ associated with $m_0^{\ast\ast}$ and
$m_0(0.2561) = 2.337\times 10^{-5}$ associated with $m_0^\ast$, such that $f(0.2560) = -2.0711 \simeq f(0.2561) = -2.0242$},
whereas for $p\geq 0.4$ it is continuous at $T_0$.
It is therefore expected that $0.2< p^\ast \leq 0.4$ for this spin number.

Interestingly, as can be verified from the inset of Fig. \ref{fig:m0_minimizing_f},
for $q=1000$ there is a concentration ($p=0.3$) such that, when cooled, the system first exhibits a continuous transition
and then a discontinuous transition at a lower temperature.
A similar behavior has been found for other concentrations.
For $q=6$ this phenomenon has not been clearly observed. Thus, it is expected that for small $q$, 
the dual transition is either a weak effect
or does not exist. 
For instance, 
the $p=0.2$ graph may obscure a small finite magnetization in
the vicinity of a second order critical point.
If such a finite magnetization exists, then $p^\ast < 0.2$.
This may exemplify the observation that in the case where in addition to the second order line,
a first order branch is launched, $p^\ast$ being the concentration where
the second order line terminates, is well defined.

\begin{figure}[ht!]
\begin{center}
\includegraphics[width = 1.\columnwidth]{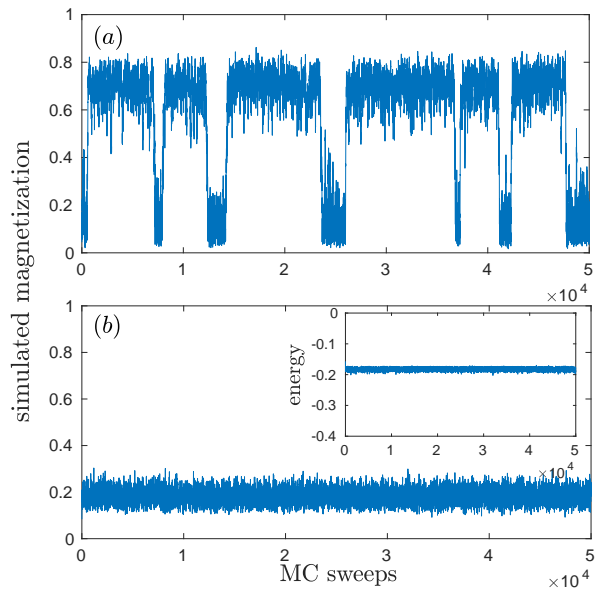}
\caption{MC time variation of the simulated MF magnetization given by \eqref{eq:SM}. The simulated system has $N = 500$ spins and a total spin number $q=6$.
For clarity, the presented observables capture a portion of $10\%$ of the total number of sweeps.
Two different concentrations corresponding to a first or a second order transition together
with temperatures in the vicinity of $T_0$ (to be precise $T = T_0 + \frac{5}{N}= T_0 + 0.01$), are chosen. (a)
$p=0.1$ and $T = 0.2661$.
(b) $p=0.6$ and $T =0.4199$.
The magnetization is averaged over $10$ different runs.
Energy per site (also averaged over $10$ different runs) against MC time is plotted in the inset.
The energy fluctuates around 
$\langle\mathrm{energy}\rangle \simeq -0.1822$ ($\mathrm{std(energy)}\simeq 0.0037$), in plausible agreement with $\varepsilon_c = -0.172011...$ 
due to \eqref{eq:e_c}.}
\label{fig:metropolis}
\end{center}
\end{figure}

We next discuss the results of Metropolis \cite{Metropolis1953} MC simulations performed to
capture numerically the concentration-dependent changeover phenomenon for the MF model.
We consider the simulated magnetization (SM)
\begin{equation}
\label{eq:SM}
{\rm{SM}} = p\frac{q_0 z_0-1}{q_0-1} + (1-p)\frac{q z-1}{q-1}\;,
\end{equation}
where $z_0$ and $z$ are the maximal fractions of monochromatic spins
in strong and weak locations, respectively, and $q_0=2$. 
A sample of $N=500$ spins is used in the simulations where the SM together with the energy (per site) proportional to \eqref{eq:H_MF} are observed
over a total number of $5\times 10^5$ MC sweeps.
In Fig. \ref{fig:metropolis}(a)
it is seen that the SM performs multiple ``flip-flops" between two values of ordered and disordered states, as expected
from a system undergoing a first order transition. On the other hand, in Fig. \ref{fig:metropolis}(b)
the system exhibits a typical second order transition by means of fluctuations around a
single value.

\begin{figure}[hbt!]
\begin{center}
\vspace*{.06cm}
\includegraphics[width = 1.\columnwidth]
{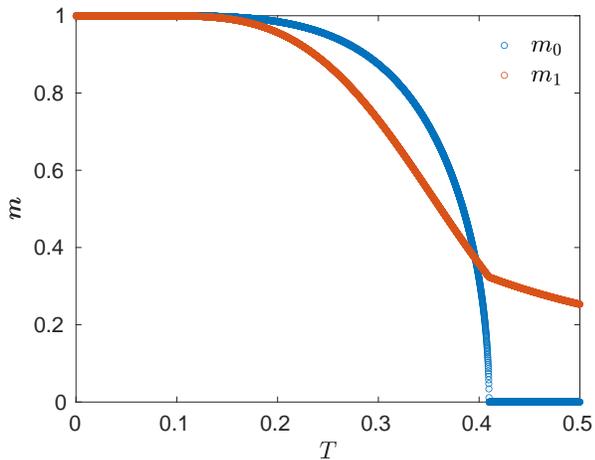}
\caption{Magnetization components minimizing \eqref{eq:betaf} as a function of temperature,
for $q=6$ and $p=0.6$. The pseudo-critical temperature is
$T_0 = 0.4099$. The pseudo-critical magnetization at $T_0$ is $m_1 = 0.3235$
(c.f. $m_1^\ast = 0.323567...$ due to \eqref{eq:m_c}).
\label{fig:m_vec_minimizing_f}}
\end{center}
\end{figure}

We conclude this section by delving into the continuous regime of Fig. \ref{fig:m0_minimizing_f},
first by solving \eqref{eq:Tc2nd_exact} taking $q=6$ and $p=0.6$, to
give $T_c=\beta_c^{-1} = 0.409809...$.
Next, from Fig. \ref{fig:m_vec_minimizing_f}, where a plot of $\boldsymbol{m}$ for $q=6$ and $p=0.6$ against temperature is presented,
it can be deduced that since the value of $T_0$ agrees well with $T_c$, $\boldsymbol{m}$ is continuous at $T_c$.
It should be emphasized at this point that, unlike $m_0$, $m_1$ is actually not a proper order parameter as it does not vanish at any finite temperature.
It has, however, a computational value. 
For instance,
it can be verified from Fig. \ref{fig:m_vec_minimizing_f} that
$m_1$ is not differentiable at $T_c$ which makes $T_c$ a unique temperature.
Furthermore, substituting $q=6,\ p=0.6,\ \beta_c = 2.440161...$ in \eqref{eq:m_c},
gives $m_1^\ast = 0.323567...$ which is in excellent agreement with $m_1$ computed at $T_0$.

\section{Summary and discussion}
\label{sec:summary}
It is demonstrated on the standard Potts model that in a mixed system of
 spins randomly colored in $q_0$ or $q>q_c\geq q_0$ colors, the transition order
depends on the concentration $p$ of $q_0$-type spins. For that system there is a critical concentration $0<p^\ast<1$ such that the transition is discontinuous when $p$
is below $p^\ast$.
For a small concentration of strong regions and a large total number of colors, a changeover phenomenon is verified analytically and numerically.
In the large $q$ limit, $p^\ast$ displays a monotonically decreasing beheavior.
This means that in the asymptotic regime, with increasing values of $q$,
it is entropically beneficial to construct 
sparser monochromatic fractals that may have a declining growth constant, than simple 
large-scale clusters. 

The Bragg-Williams approximation, applied to the MF model,
uncovers the presence of the marginal concentration for that model,
where the strong regions are occupied by Ising-like spins.
Unlike in the standard model where it is argued that a first order transition occurs for concentrations below $p^\ast$,
in the MF model we conclude the presence of $p^\ast$ from the other direction,
that is, the second order critical
temperature, as a function of $p$, is 
defined on some interval with a left endpoint which is equal to $p^\ast$.

It is inspected that for $q$ large
there are concentrations hosting both continuous and discontinuous
transitions. A similar phenomenon, namely, a sequential continuous and discontinuous transition
occuring at some values of $r$, has been reported for the IC model \cite{Krasnytska2016}
where the Bragg-Williams approximation in the context of the MF model was employed.

A reasonable explanation for the large $q$ dual transition
is that, at the higher temperature it is
entropically favorable to typically leave the weak
regions disordered, hence, Ising-like long range
order is established in the strong regions. However, when the temperature is lowered,
at some point it becomes energetically beneficial to
expand the predetermined LRO by invading the weak regions.
It is yet not clear whether the dual transition is manifested for any $q>2$.

The simulations produce observables
that have a different qualitative behavior for different concentrations.
In the standard model case the PC energy PDF changes its
shape from dual to single peaked, with increasing values of $p$
and in the MF case the magnetization has a (noisy) typical single versus non-single
valued temporal behavior, for different values of $p$.
Thus, in both cases, the essence of $p^\ast$ as a changeover point is revealed.

Our model provides a rather general framework of governing
the transition order of well known Potts systems, without 
interfering with the interaction content of such systems
when no heterogeneity in the number of colors is introduced, in the sense that a
changeover phenomenon is observed for \textit{any} setup of
strong and weak colors.
We believe that the HP machinery can be applied to other Potts systems
such that a concentration-dependent changeover
phenomenon is detected in those systems.

%


\end{document}